\documentclass[prd,aps,showpacs,twocolumn,eqsecnum,nofootinbib]{revtex4-1}
\usepackage{amsmath}
\usepackage{graphicx}
\usepackage{latexsym}
\usepackage{amssymb}
\usepackage{color}
\usepackage{bm}
\usepackage{amssymb}
\usepackage{enumitem}
\usepackage{hyperref}
\usepackage{subfigure}
\usepackage{array}
\usepackage[english]{babel}
\usepackage{tensor}
\usepackage{natbib}
\allowdisplaybreaks
\usepackage{ulem}

\def\be{\begin{equation}}
\def\ee{\end{equation}}
\def\ba{\begin{eqnarray}}
\def\ea{\end{eqnarray}}
\def\l{\left}
\def\r{\right}

\begin{document}

\title{Constraining cosmological scaling solutions of a Galileon field}
\author{In\^es S.~Albuquerque$^1$, Noemi Frusciante$^1$, Matteo Martinelli$^{2,3}$}

\affiliation{\smallskip\smallskip$^1$Instituto de Astrof\'isica e Ci\^encias do Espa\c{c}o, 
Faculdade de Ci\^encias da Universidade de Lisboa,  Campo Grande, PT1749-016 Lisboa, Portugal\\
\smallskip
$^2$ INAF - Osservatorio Astronomico di Roma, via Frascati 33, 00040 Monteporzio Catone (Roma), Italy\\
\smallskip
$^3$ Instituto de F\'isica T\'eorica UAM-CSIC, Campus de Cantoblanco, 28049 Madrid, Spain
}

\begin{abstract}

We study a Lagrangian with a cubic Galileon term and a standard scalar-field kinetic contribution with two exponential potentials. In this model the Galileon field  generates scaling solutions in which the density of the scalar field $\phi$  scales in the same manner as the matter density at early-time. These solutions  are of high interest because the scalar field can then be compatible with the energy scale of particle physics and can alleviate the coincidence problem. The phenomenology of linear perturbations is thoroughly discussed, including all the relevant effects on the observables. Additionally, we use cosmic microwave background  temperature-temperature and lensing power spectra by Planck 2018, the baryon acoustic oscillations  measurements from the 6dF galaxy survey and SDSS  and supernovae type Ia  data from Pantheon in order to place constraints on the  parameters of the model. We find that despite its interesting phenomenology, the model we investigate does not produce a better fit to data with respect to $\Lambda$CDM, and it does not seem to be able to ease the tension between high and low redshift data.

\end{abstract}

\maketitle

\section{Introduction}

The standard cosmological scenario, or  $\Lambda$-cold-dark-matter ($\Lambda$CDM) model, is the most accepted model to explain the observable Universe and its late-time accelerated expansion. However, some mild observational tensions among different datasets emerge in this scenario, namely Planck data analyzed within the $\Lambda$CDM model \cite{Planck:2018vyg} show a tension between 4 and 6$\sigma$ with late-time, model-independent measurements \cite{Riess:2019cxk,Wong:2019kwg,BOSS:2014hwf,BOSS:2012dmf,SDSS:2008tqn,Freedman:2019jwv,Yuan:2019npk,Riess:2021jrx} of the  Hubble constant $H_0$ and with weak lensing data for the estimation of the present time amplitude of the matter power spectrum in terms of $\sigma_{8,0}$ \cite{deJong:2015wca}, see also Refs.~\cite{DiValentino:2020vvd,DiValentino:2020zio,DiValentino:2021izs,DiValentino:2018gcu} for general overviews. We can further mention the Planck lensing anomalies about the excess of lensing in the temperature power spectrum \cite{Planck:2018vyg}. Additionally, the cosmological constant, $\Lambda$, responsible for the late-time accelerated expansion, is plagued by theoretical issues, such as (among others) the well known Cosmological Constant Problem \cite{Weinberg:1988cp,Carroll:2000fy,Weinberg:2000yb,Padilla:2015aaa}, i.e.\ the discrepancy between the theoretically predicted value for $\Lambda$ and the observed one, and by the Coincidence Problem \cite{Velten:2014nra}, namely why the magnitude of the energy densities for matter and the cosmological constant are comparable today. These issues are motivating the search for new physics beyond the $\Lambda$CDM model \cite{CANTATA:2021ktz}.

In this paper we specialize on a modified gravity (MG) theory, the Galileon one \cite{Nicolis:2008in,Deffayet:2009mn,Deffayet:2011gz}, whose generalization is equivalent \cite{Kobayashi:2010cm} to the  Horndeski theory \cite{Horndeski:1974wa}. For a long time Horndeski theory has been considered to be constructed from the most general action for a scalar field coupled to gravity that leads to second order equations of motion, but later it has been found that more general actions can be constructed, namely those of the beyond Horndeski \cite{Gleyzes:2014dya} and DHOST \cite{Langlois:2015cwa} theories. The Galileon/Horndeski theory allows for self-accelerating solutions which have been the basis of inflationary scenarios \cite{Kobayashi:2010cm,Burrage:2010cu,Mironov:2019qjt,Creminelli:2010qf,Renaux-Petel:2011rmu,Kamada:2010qe,DeFelice:2013ar,Kobayashi:2011pc,Gao:2011qe,Takamizu:2013gy,Frusciante:2013haa} and late-time explanations of cosmic acceleration \cite{Deffayet:2010qz,Charmousis:2011bf}. The observation of the Gravitational Waves (GW) event GW170817 \cite{LIGOScientific:2017vwq} and of its electromagnetic counterpart GRB170817A \cite{Goldstein:2017mmi}, set a stringent bound on the speed of propagation of GWs \cite{LIGOScientific:2017zic}, which in turn severely constrains the form of the Galileon action \cite{Creminelli:2017sry,Ezquiaga:2017ekz,Baker:2017hug,Amendola:2017orw}, leaving only a sub-class still viable. Within the surviving Galileon models, data analysis with Planck data alone found that $H_0$ is consistent with its local determination respectively at 1$\sigma$ for the generalized cubic covariant Galileon model \cite{Frusciante:2019puu} and at 2$\sigma$ for the Galileon ghost condensate \cite{Peirone:2019aua}, resolving the $H_0$ tension. For the latter, a joint data analysis of cosmic microwave background (CMB) radiation, baryonic acoustic oscillations (BAO), supernovae type Ia (SN) and redshift-space distortions (RSD) showed that it is also statistically preferred over the $\Lambda$CDM scenario due to  suppressed large-scale temperature anisotropies and a peculiar behavior of the scalar field equation of state in the early-time expansion history \cite{Peirone:2019aua}.  
 
In this work, we are interested in a particular class of Galileon models, specifically the one in which the scalar field can give rise to \textsl{scaling solutions}. Scaling solutions \cite{Copeland:1997et,Ferreira:1997hj,Liddle:1998xm,Barreiro:1999zs,Guo:2003rs,Guo:2003eu,Tsujikawa:2004dp,Piazza:2004df,Pettorino:2005pv,Amendola:2006qi,Ohashi:2009xw,Gomes:2013ema,Chiba:2014sda,Amendola:2014kwa,Albuquerque:2018ymr,Frusciante:2018aew,Amendola:2018ltt} are characterized by a constant ratio between the energy density of the matter components and that of the scalar field. In this case, the density of the scalar field is not negligible compared to the other components even at early time, thus the model shows compatibility with the energy scale associated with particle physics. This feature might alleviate the Coincidence Problem since the initial conditions in the scaling regime are fixed by the model parameters.  

A general Galileon Lagrangian allowing for scaling solutions has the form $L = Xg_2(Y)-g_3(Y)\Box \phi$ \cite{Frusciante:2018aew}, where $X \equiv - \partial_{\mu} \phi \partial^{\mu} \phi / 2$ and $g_2$, $g_3$ are general functions of  $Y=X e^{\lambda \phi} $ with $\lambda$ being a constant. For this Lagrangian, scaling solutions are also present when the scalar field $\phi$ is coupled to non-relativistic matter with a constant coupling  $Q$ \cite{Frusciante:2018aew}. Concrete models have been proposed for $g_3$, such as $g_3 = a_1 Y + a_2 Y^2$, with $a_1$ and $a_2$ constants, together with an exponential potential and a direct coupling between the scalar field and matter \cite{Gomes:2013ema}, or $g_3 = A \ln Y$, with $A$ being a constant, together with a standard kinetic term and two exponential potentials \cite{Albuquerque:2018ymr}. For the resulting model, the density associated to the $g_3$ term gives important contributions to the field density during scaling radiation and matter eras, but then it is subdominant at later-time relative to the density associated to the standard field Lagrangian characterizing $g_2$.
This feature is  expected to accommodate  the observational data of galaxy and Integrated-Sachs-Wolfe (ISW) cross-correlations, which indeed do not seem to prefer  dominance  of cubic interactions at late-time \cite{Kimura:2011td,Renk:2017rzu,Kable:2021yws}. Furthermore,
the $g_3$ term modifies the gravitational couplings felt by matter and light, leading to a modified evolution of perturbations \cite{Albuquerque:2018ymr}. These signatures can be used to distinguish the model from standard Quintessence and $\Lambda$CDM \cite{Banerjee:2020xcn}. In this work we will investigate this model, to which we will refer to as the Scaling Cubic Galileon (SCG) model. We will provide a thorough analysis of cosmological perturbations and their effects on CMB anisotropies, lensing potential and growth of structures and finally we will present the cosmological constraints obtained using  combinations of data sets. 

This paper is organized as follows. In Sec.~\ref{sec:II} we review the theoretical framework of the  SCG model. In  Sec.~\ref{Sec:CosmoObs} we present the theoretical predictions for some cosmological observables.  In Sec.~\ref{Sec:Constraints} we provide the cosmological constraints using the Markov chain Monte Carlo (MCMC) method. Finally we conclude in Sec.~\ref{Sec:Conclusion}.

\section{The Scaling Cubic Galileon Model} \label{sec:II}

In this section we review the SCG  model, whose action reads:
\be
\mathcal{S}=\int d^4x\sqrt{-g}\l[\frac{R}{16 \pi G_N}+\mathcal{L}_{\rm SCG}(\phi,X)+\mathcal{L}_{ \gamma}(g_{\mu\nu},\chi_{\gamma})\r]\,,
\ee
where $G_N$ is the Newtonian gravitational constant, $g_{\mu\nu}$ is the metric and $g$ is its determinant, $\mathcal{L}_{\gamma}$ is the Lagrangian of matter fluids, $\chi_{\gamma}$, and $\mathcal{L}_{\rm SCG}$ is the  Lagrangian describing the SCG model, defined as follows \cite{Albuquerque:2018ymr}
\begin{equation}
\mathcal{L}_{\rm SCG}=  X-V_1 e^{-\beta_1 \phi} - V_2 e^{-\beta_2 \phi} - A \ln Y \Box \phi , \label{eq:SCGaction}
\end{equation}
with  $V_1$, $V_2$, $\beta_1$,  $\beta_2$, $A$ and $\lambda$ being constants. $V_1$, $V_2$, $\beta_1$ and $\beta_2$ are positive defined and $\beta_1$ and $\beta_2$ are chosen such that $\beta_1 \gg \mathcal{O} (1)$  in order to satisfy early-time constraints on the field density parameter from Big Bang Nucleosynthesis and  CMB
measurements \cite{Copeland:1997et,Albuquerque:2018ymr}, while $\beta_2 \lesssim \mathcal{O} (1)$ in order to realize the late-time acceleration \cite{Barreiro:1999zs}. The double exponential potential is chosen because it provides the necessary mechanism for the scalar field to exit the early-time scaling regime during which its density is proportional to that of the matter components (hence the name SCG) into a late-time epoch of cosmic acceleration \cite{Albuquerque:2018ymr} (another example in the context of standard Quintessence is Ref.~\cite{ Barreiro:1999zs}). In the following we adopt the unit $8 \pi G_N = 1$.

\subsection{Background Equations}
Let us consider a flat Friedmann-Lemaître-Robertson-Walker (FLRW) background described by the line element
\begin{equation}
    ds^2 = - dt^2 + a^2 (t) \delta_{ij} dx^i dx^j , 
\end{equation}
where $a(t)$ is a time-dependent scale factor. The modified Friedmann equations for the SCG model read
\begin{align}
    3H^2 &= \rho_{\gamma} + \rho_{\phi} , \label{eq:FriedConst}  \\
    2\Dot{H} + 3H^2 &= - \left( p_{\gamma} + p_{\phi} \right) , \label{eq:AccEq}
\end{align}
where $H(t) \equiv \Dot{a}/a$ is the Hubble expansion rate and a dot represents a derivative with respect to cosmic time. The quantities $\rho_{\gamma}$ and $p_{\gamma}$ correspond, respectively, to the energy density and pressure of the standard matter fluids, namely, cold dark matter (c), baryons (b) and photons (r), and are related through the barotropic equation of state $p_{\gamma} = (\gamma - 1) \rho_{\gamma}$.  The constant barotropic coefficient is $\gamma_{\rm b,c} = 1$ for baryons and cold dark matter and $\gamma_{\rm r} = 4/3$ for photons. Additionally, $\rho_{\phi}$ and $p_{\phi}$ are the energy density and pressure of the scalar field $\phi$ defined as
\begin{align}
    &\rho_{\phi} = \frac{\Dot{\phi}^2}{2} + V_1 e^{-\beta_1 \phi} + V_2 e^{- \beta_2 \phi} + 6 H A \Dot{\phi} - A \lambda \Dot{\phi}^2 , \\
    &p_{\phi} =  \frac{\Dot{\phi}^2}{2} - V_1 e^{- \beta_1 \phi} - V_2 e^{- \beta_2 \phi} - 2 A \left( \frac{\lambda \Dot{\phi}^2}{2} + \Ddot{\phi} \right)\, . \end{align}
Finally, the equation of motion for the scalar field is:
\begin{equation} \label{eq:EqMotSF}
    \frac{1}{a^3} \frac{d}{dt} \left( a^3 J \right) = P \,,
\end{equation}
with
\begin{align}
&J= \Dot{\phi} + 6 A H - 2 A \lambda \Dot{\phi}\,,\\
&P=\beta_1 V_1 e^{- \beta_1 \phi} + \beta_2 V_2 e^{-\beta_2 \phi}\,.
\end{align}

It is possible to  rewrite the dynamics of the background evolution of the SCG in terms of the following dimensionless variables:
\begin{equation} \label{eq:SCGVar}
    x = \frac{\Dot{\phi}}{\sqrt{6} H} , \ \  y_1 = \frac{\sqrt{V_1 e^{-\beta_1 \phi}}}{\sqrt{3} H} , \ \ y_2 = \frac{\sqrt{V_2 e^{-\beta_2 \phi}}}{\sqrt{3} H} ,
\end{equation}
together with
\begin{equation} \label{eq:Densi}
    \Omega_{\rm m} = \frac{\rho_{\rm m}}{3H^2} , \ \ \ \Omega_{\rm r} = \frac{\rho_{\rm r}}{3H^2} \ \ \ \text{and} \ \ \ \Omega_{\phi} = \frac{\rho_{\phi}}{3H^2} \,,
\end{equation}
where $\mbox{m=c+b}$.
Eqs.~\eqref{eq:FriedConst}, \eqref{eq:AccEq} and \eqref{eq:EqMotSF} can then be rearranged into an autonomous system of first-order differential equations:
\begin{align}
    x' &= \frac{1}{\sqrt{6}} f(x,y_1,y_2) - \tilde{h}(x,y_1,y_2) x , \label{eq:SCGDS1} \\
    y_{i}' &= - \sqrt{\frac{3}{2}} \beta_{i} x y_{i} - \tilde{h}(x,y_1,y_2) y_{i} , \label{eq:SCGDS2} \\
    \Omega_{r}' &= - 4 \Omega_r- 2 \tilde{h}(x,y_1,y_2) \Omega_r , \label{eq:SCGDS3}
\end{align}
where a prime denotes a derivative with respect to $\ln a$, $i=1,2$ and we have defined the functions $f \equiv \Ddot{\phi}/H^2$ and $\tilde{h} \equiv \Dot{H}/H^2$. The latter two can be given completely in terms of the variables introduced in eqs.~\eqref{eq:SCGVar} and \eqref{eq:Densi}. Since their expressions are quite long, we refer the reader to Ref.~\cite{Albuquerque:2018ymr} for their explicit forms.  Finally, the Friedmann constraint given by eq.~\eqref{eq:FriedConst} becomes
\begin{equation} \label{eq:SCGFriedConst}
    \Omega_{\rm m}+ \Omega_{\rm r} + \Omega_{\phi} = 1  , 
\end{equation}
where the scalar field density parameter $\Omega_{\phi}$ reads
\begin{equation} \label{eq:SCGOmegaPhi}
    \Omega_{\phi} = x^2 + y_1^2 + y_2^2 + 2 x A \left( \sqrt{6} - \lambda x \right) . 
\end{equation}

As such, the SCG model is left with a set of four extra free parameters with respect to $\Lambda$CDM: $\{ A, \beta_1, \beta_2 , \lambda \}$. The range of possible values for these parameters is constrained by the enforcement of theoretical viability conditions for the evolution of both the background and the perturbation sector, including, for example, the conditions for the absence of ghost ($Q_s>0$) and gradient ($c_s^2>0$)  instabilities, which respectively read \cite{Albuquerque:2018ymr}:
\ba
Q_s&=&\frac{3 x^2 \left(1-2 A \lambda+6 A^2 \right)}
{\left(1-\sqrt{6} A x\right)^2} \,,  \\
c_s^2&=&\frac{3x(1-2A\lambda-2A^2)+4 \sqrt{6} A}
{3x \left( 1-2 A \lambda+6 A^2\right)}  \,.
\ea

We can note that in the limit $A\rightarrow 0$ the above conditions are fully satisfied because the model reduces to  standard Quintessence; if $A\neq0$ these translate into theoretical bounds on $A$ and $\lambda$. A full discussion of their impact on the parameter space can be found in Ref.~\cite{Albuquerque:2018ymr}. 

Under the validity of the above conditions, the SCG model has very interesting features: it is possible to realize scaling solutions followed by the dark energy attractor, in this case since the scaling fixed point is stable a second potential is necessary to exit this regime and to start the late-time accelerated expansion; during the early phase of the Universe the scalar field density arising form the cubic $g_3$ term dominates over the standard $g_2$ term while the opposite happens at late-time;  having $A\neq 0$ allows for a wider parameter space for $\beta_1$ and $\beta_2$ compared to a standard Quintessence model, in particular it allows for $\beta_2^2>2$; the scalar field equation of state today, $w_\phi\equiv p_\phi/\rho_\phi$, can be closer to $-1$ than in standard Quintessence. 

Finally let us stress that in order to solve the system it is still necessary to provide a set of initial conditions (ICs). In particular, since the model is characterized by a scaling fixed point we can use such scaling solution at early-time to fix the ICs through the model parameters. The ICs for $x_i$ and $y_{1,i}$ are then selected to correspond to the radiation scaling critical point, given by \cite{Albuquerque:2018ymr}
\begin{equation}
    \left( x_i, y_{1,i} \right) = \left( \frac{2 \sqrt{6}}{3 \beta_1} , \frac{\sqrt{12 + 6 A ( 3 \beta_1 - 4 \lambda) }}{3 \beta_1} \right) ,
\end{equation}
while the IC for $y_{2,i}$ is then determined by iteratively solving the background equations until the Friedmann constraint \eqref{eq:SCGFriedConst} is satisfied at present time.

\subsection{Linear Perturbations} \label{sec:IIB}

Let us also consider the perturbed FLRW line element in Newtonian gauge given by
\begin{equation}
    ds^2 = - \left( 1 + 2 \Psi \right) dt^2 + a^2(t) \left( 1 - 2 \Phi  \right) \delta_{ij} dx^i dx^j , 
\end{equation}
where $\Psi(t,x_i)$ and $\Phi(t,x_i)$ correspond to the gravitational potentials and obey the Poisson and lensing equations, given  in Fourier space as follows \cite{Bean:2010zq,Silvestri:2013ne,Pogosian:2010tj}
\begin{align}
    - \frac{k^2}{a^2} \Psi &= 4 \pi G_N \mu (t, k) \rho \Delta , \label{eq:Poisson} \\
    - \frac{k^2}{a^2} \left( \Phi + \Psi \right) &= 8 \pi G_N \Sigma (t, k) \rho \Delta ,
\end{align}
where $k$ is the comoving wavenumber and the comoving density contrast is defined as $\Delta_i \equiv \delta_i + 3H v_i/k $ with $\delta_i = \delta \rho_i / \rho_i $ the density contrast, $\rho_i(t)$ the background energy density of a matter component $i$ and $v_i$ the irrotational component of the peculiar velocity. The two phenomenological functions $\mu$ and $\Sigma$  are, respectively, the effective gravitational coupling, defining the deviation with respect to $\Lambda$CDM on the clustering of
matter and the light deflection parameter characterizing the modifications introduced in the paths travelled by photons on cosmological scales via the modification of the lensing potential $\Phi + \Psi$. Lastly, the difference between the gravitational potentials can be described by the gravitational slip parameter $\eta= \Phi/\Psi$. The $\Lambda$CDM limit is obtained for $\mu = \Sigma = \eta = 1$. As such, any departure from unity translates a signature of modified gravity.

One can obtain approximate analytical expressions for $\mu$, $\Sigma$ and $\eta$ by assuming the quasi-static approximation (QSA)\cite{Boisseau:2000pr,DeFelice:2011hq}, which, in the case of Galileon theory, has been proven to be a valid assumption for modes with $k \gtrsim 0.001 $ h/Mpc \cite{Pogosian:2016pwr,Frusciante:2019xia}. In the specific case of the SCG model, the QSA yields \cite{Albuquerque:2018ymr}
\begin{equation} \label{eq:SCGMu}
\mu = \Sigma = 1 + \frac{6 A^2 x^2 }{Q_sc_s^2(1- \sqrt{6} A x)^2}\,,  \quad \eta = 1. 
\end{equation}
Therefore the presence of the cubic coupling  introduces modifications on both the growth of structures $( \mu \neq 1 )$ and the evolution of the weak lensing potential $( \Sigma \neq 1 )$ for all $A \neq 0$. Additionally,  $\mu = \Sigma > 1$ for all viable values of the parameters. As a consequence, we expect the gravitational interaction to be stronger than the one of $\Lambda$CDM.  

\begin{table}[t!]
\begin{center}
\begin{tabular}{|c|c|c|c|c|c|c|}
\hline
Model & $\beta_1$ & $\beta_2$ & $A$ & $ \lambda$ & 
$w_{\phi}^{(0)}$\\ 
\hline\hline
M1  & 100 & 0.7 & -0.3 & 154  & -0.993 \\ \hline 
M2  & 100 & 0.7 & 0.09 & -8   &  -0.988 \\ \hline 
M3  & 100 & 0.7 &-0.28 & 148.3& -0.993 \\ \hline
M4  & 100 & 2.5 & -1   &  150 & -0.975 \\   \hline
\end{tabular}
\end{center}
\caption{\label{tab:Models}
Model parameters $\{ \beta_1 , \beta_2 , A , \lambda \}$ for the four case-study SCG models. In addition to the parameter values, we also show today's dark energy equation 
of state $w_{\phi}^{(0)}$.}
\end{table}

Let us stress that while here we present the equations under the QSA, in what follows we shall not rely on this approximation and instead evolve the full linear perturbation equations. The discussion in this section will, nevertheless, help in the interpretation of the results we present in Sec.~\ref{sec:CosmPert}.

\section{Cosmological Observables}\label{Sec:CosmoObs}

\subsection{Methodology} \label{Sec:Method}

The main goal of this project is to investigate the evolution of linear cosmological perturbations and obtain observational constraints on the SCG model. To achieve this, we make use of the Effective Field Theory (EFT) formalism \cite{Gubitosi:2012hu,Bloomfield:2012ff}  (see Ref.~\cite{Frusciante:2019xia} for a review) and then resort to the public available Einstein-Boltzmann code \texttt{EFTCAMB} \cite{Hu:2013twa}\footnote{Web page: \url{http://www.eftcamb.org}} and MCMC engine \texttt{EFTCosmoMC} \cite{Raveri:2014cka}. As such, in this section we briefly discuss the implementation of the SCG model in \texttt{EFTCAMB}.

The EFT offers a model-independent framework to describe both the background evolution of the Universe and the behaviour of linear cosmological perturbations in gravity theories with a single additional scalar DoF. This description is made in terms of a set of free functions of time known as EFT functions. The EFT action encompassing Galileon theory with the additional consideration of the bound on the speed of propagation of tensor modes coming from GWs ($c_T^2=1$) is
\begin{align}
    \mathcal{S} = &\int d^4 x \sqrt{-g} \frac{m_0^2}{2} \Bigg\{ \left( 1 + \Omega (t) \right) + \frac{2 \Lambda (t)}{m_0^2} - \frac{2 c(t)}{m_0^2} \delta g^{00} \nonumber \\
    &+ H_0^2 \gamma_1 (t) \left( \delta g^{00} \right)^2 - H_0 \gamma_2 (t) \delta g^{00} \delta K \Bigg\} + \mathcal{S}_{\gamma} \ , \label{eq:EFTAction}
\end{align}
where $m_0^2$ is the Planck mass and $\delta g^{00}$ and $\delta K$ are the linear perturbations of the upper time-time metric component and the trace of the extrinsic curvature $K^{\mu}_{\nu}$, respectively. The time dependent prefactors  $\{ \Omega, \Lambda, c, \gamma_1, \gamma_2 \}$ are the aforementioned  EFT functions.  

The action \eqref{eq:EFTAction} can then be connected to a specific Galileon model by finding the corresponding forms of the EFT functions through a procedure known as \textit{mapping} \cite{Gleyzes:2013ooa,Bloomfield:2013efa,Frusciante:2016xoj}. For the SCG model, we follow the procedure depicted in Ref.~\cite{Frusciante:2016xoj} and find:
\begin{align}
    \frac{c a^2}{m_0^2} &= 3 \mathcal{H}^2 x^2 + A \mathcal{H}^2 \Big[ 3 \sqrt{6} x - 6 \lambda x^2 - \sqrt{6} b(x)  \Big] , \\
    \frac{\Lambda a^2}{m_0^2} &= 3 \mathcal{H}^2 \left( x^2 - y_1^2 - y_2^2 \right) - 2 \mathcal{H}^2 A \Big[ 3 \lambda x^2 + \sqrt{6} b(x) \Big] , \\
    \gamma_1 &= \sqrt{\frac{3}{4}} \frac{A \mathcal{H}^2}{a^2 H_0^2} \left( b(x) - 3x \right) , \label{eq:SCGG1} \\ 
    \gamma_2 &= - \frac{2 \sqrt{6} A \mathcal{H}}{a H_0} x , \label{eq:SCGG2} 
\end{align}
where  $b(x) = x' + \tilde{h}x$ and $\mathcal{H}$ is the Hubble function in conformal time $\tau$. The use of conformal time and the redefinitions of $c$ and $\Lambda$ are done to match the notation of the \texttt{EFTCAMB} code. Let us note that $\Omega = 0$ in the SCG.

From the forms of $\gamma_1$ and $\gamma_2$, the two EFT functions which impact the linear perturbation sector only, we can expect modifications in the growth of perturbations to arise when the parameter $A$, present in both \eqref{eq:SCGG1} and \eqref{eq:SCGG2}, is non-zero. On the other hand, the $\lambda$ parameter, which appears on the background EFT functions $c$ and $\Lambda$, will consequently propagate its effect through the background evolution.

Finally,
we have implemented the previous EFT functions together with a background solver, composed by eqs.~\eqref{eq:SCGDS1}-\eqref{eq:SCGDS2}, in \texttt{EFTCAMB}, with the purpose of obtaining theoretical predictions for  cosmological observables and observational constraints on the SCG parameters using \texttt{EFTCosmoMC}. We stress that \texttt{EFTCAMB} solves the full linear perturbation equations without resorting to the QSA approximation.

\begin{figure}[t!]
\centering
\includegraphics[scale=0.40]{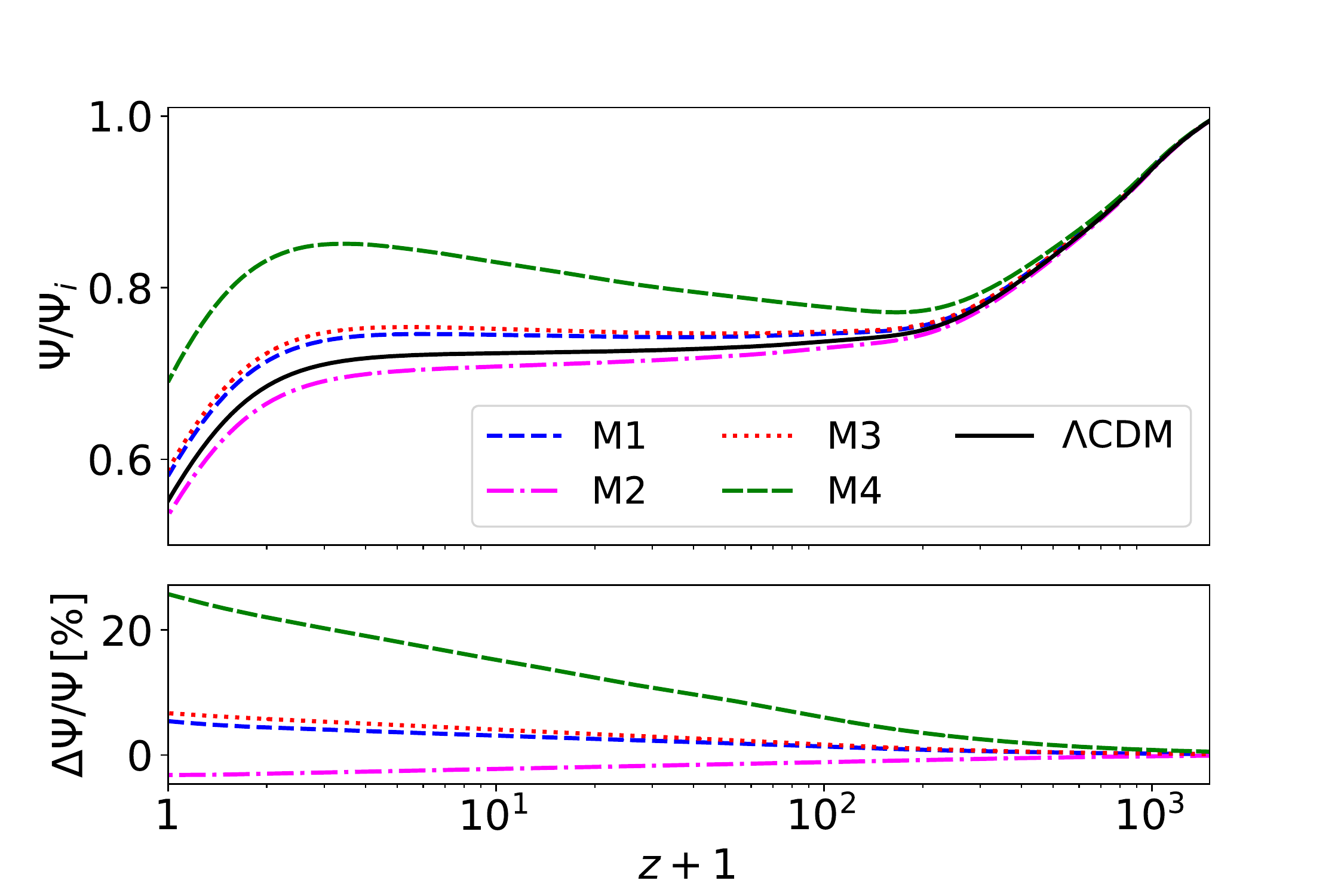} 
\caption[Evolution of $\Psi$ as a function of redshift for the M models]{(Top) Evolution of the metric potential $\Psi$ normalized by its initial value $\Psi_i$ as a function of redshift $z+1$ for the wavenumber $k = 0.01$ Mpc$^{-1}$. The evolution of $\Psi/\Psi_i$ is shown for the four models of Table \ref{tab:Models}, specifically: M1 (dashed blue line), M2 (dot-dashed magenta line), M3 (dotted red line) and M4 (long dashed green line), as well as for $\Lambda$CDM (solid black line). (Bottom) Percentage relative difference of the models' $\Psi$ with respect to $\Lambda$CDM.  \label{fig:SCGPhi}}
\end{figure}
\subsection{Phenomenology} \label{sec:CosmPert}

In this section we study the dynamics of scalar cosmological perturbations in the SCG model and analyse their impact on some cosmological observables. For our investigation, we consider four Models (M) specified by the parameters listed in Table \ref{tab:Models}. The choice of these values is purely illustrative, nevertheless they satisfy the stability requirements previously discussed. Notably we have verified that the $\beta_1$ and $\beta_2$ parameters do not have any direct impact on the phenomenology of the SCG we are going to present. Even so, they indirectly impact the parameter space, e.g.\ larger values of $\beta_2$ allow to explore regions of the parameter space with larger values of $A$ (see Ref.~\cite{Albuquerque:2018ymr}). The cosmological parameters are fixed to be: $H_0=70$ km/s/Mpc, the baryon and cold dark matter energy densities are $\Omega_{\rm b}h^2= 0.0226$ and $\Omega_{\rm c}h^2= 0.112$, where $h=H_0/100$, and finally the amplitude and tilt of the primordial power spectrum are $A_{\rm s}= 2.1 \times 10^{-9}$ and $n_s=0.96$. Let us stress that the cosmological parameters are kept fixed only for the purpose of the phenomenological analysis, as in this case we are able to trace back to the impact of MG on cosmological observables when compared to $\Lambda$CDM. We divide our discussion considering the three main effects we identify, i.e.\ on the gravitational lensing potential and its time derivative (ISW effect) and on the evolution of matter density perturbations.

\subsubsection{Impact on gravitational lensing} \label{Sec:GravLen}

\begin{figure}[t!]
\centering
\includegraphics[scale=0.40]{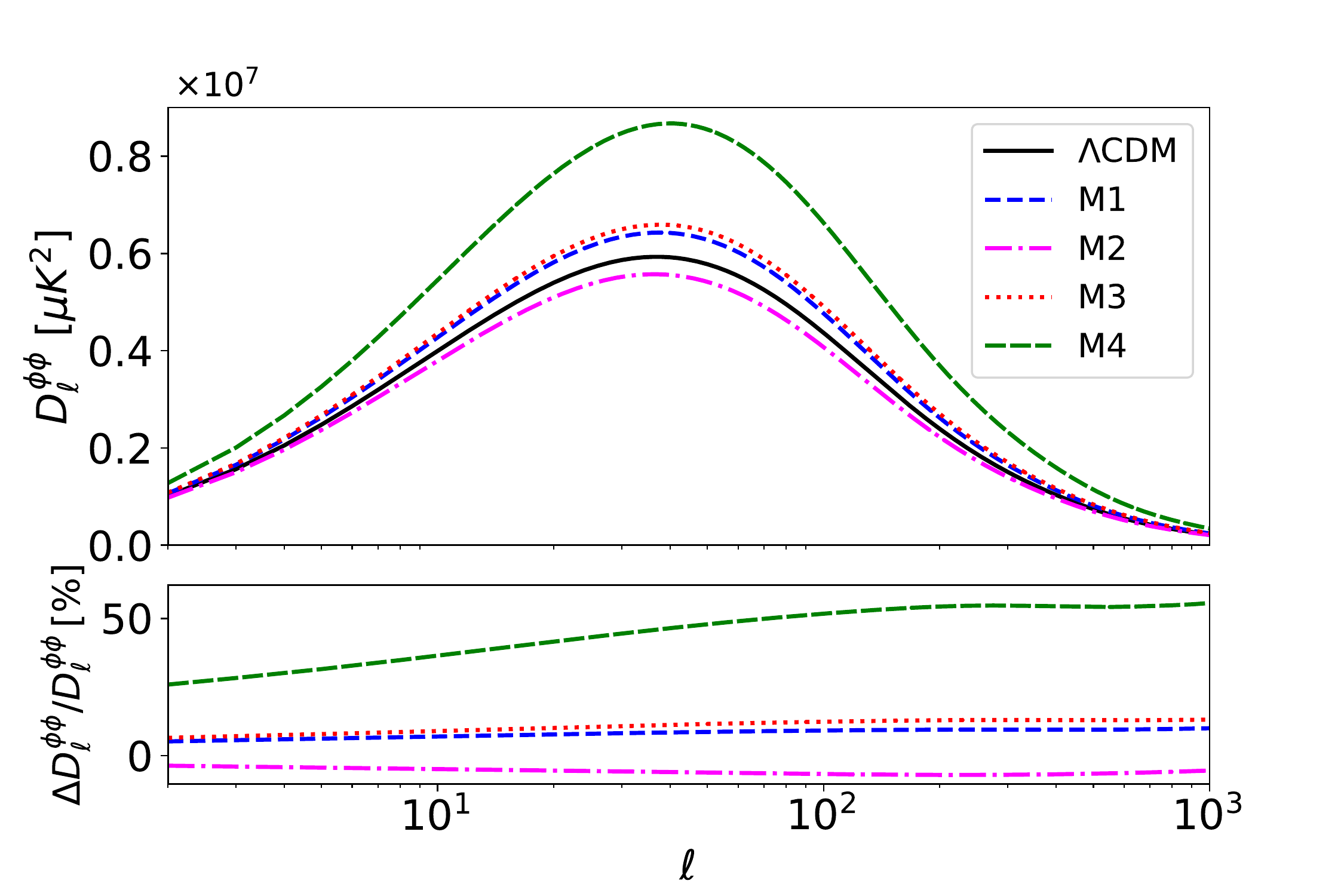} 
\caption[ISW source for the M models]{(Top) Lensing angular power spectra $D_{\ell}^{\phi \phi} = \ell (\ell + 1) C_{\ell}^{\phi \phi}/(2\pi)$ for $\Lambda$CDM (solid black line) and the four M models of Table \ref{tab:Models}, namely: M1 (dashed blue line), M2 (dot-dashed magenta line), M3 (dotted red line) and M4 (long dashed green line). (Bottom) Difference between the lensing power spectra of each of the M models and that of $\Lambda$CDM. \label{fig:SCGLenPS}}
\end{figure}

\begin{figure}[t!]
\centering
\includegraphics[scale=0.40]{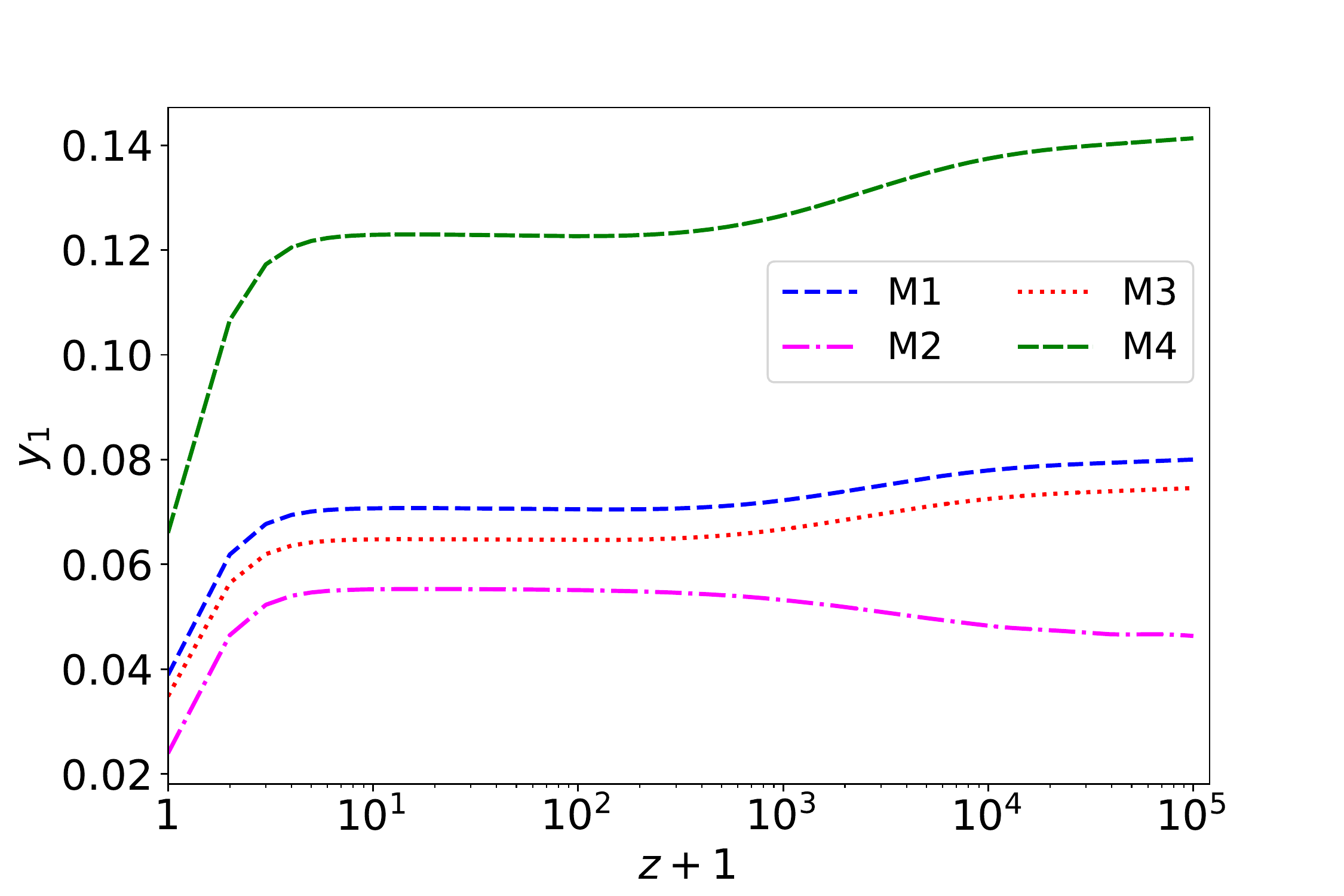} 
\caption[Evolution of y1]{Evolution of the $y_1$ variable for the models M1 (dashed blue line), M2 (dot-dashed magenta line), M3 (dotted red line) and M4 (long dashed green line).  \label{fig:SCGy1}}
\end{figure}

\begin{figure*}[t!]
\centering
\includegraphics[scale=0.35]{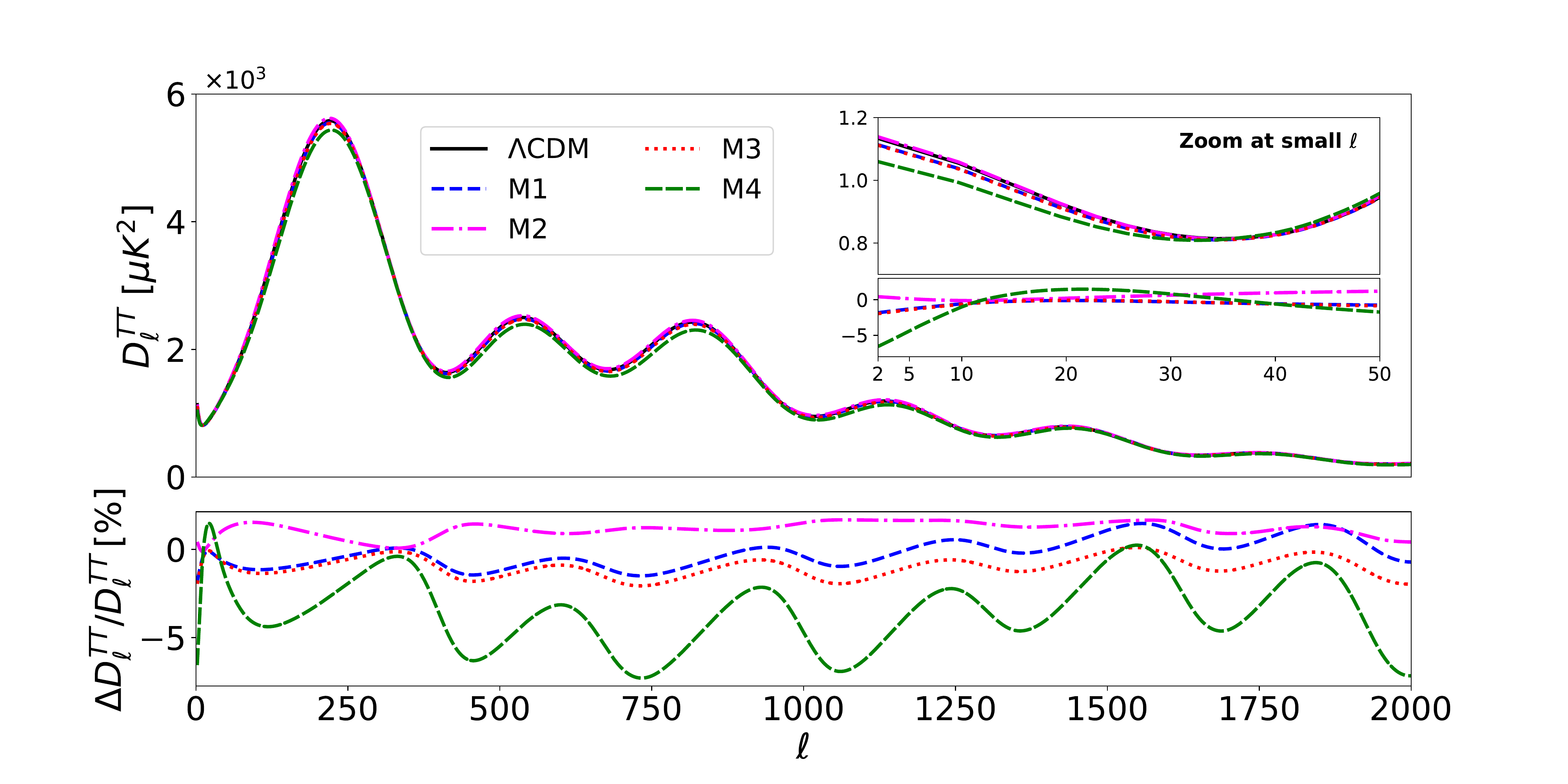} 
\caption[CMB TT power spectra for the M models]{(Top) CMB temperature-temperature (TT) angular power spectra $D_{\ell}^{TT} = \ell (\ell + 1) C_{\ell}^{TT}/(2\pi) $ for $\Lambda$CDM (solid black line) and the four M models of Table \ref{tab:Models}, namely: M1 (dashed blue line), M2 (dot-dashed magenta line), M3 (dotted red line) and M4 (long dashed green line). On the top-right corner is shown a zoom at large angular scales (small $\ell$) of this same quantity. (Bottom) Percentage relative difference between the lensing power spectra of each of the M models and that of the M models' TT power spectra with respect to $\Lambda$CDM. \label{fig:SCGCMBTTPS}}
\end{figure*}

The lensing potential $\phi_{\rm len} = \left( \Phi + \Psi \right)/2$ for the SCG model is modified compared to $\Lambda$CDM. In Figure \ref{fig:SCGPhi} we show the evolution of the gravitational potential $\Psi$ normalized by its initial value as a function of the scale factor for a fixed $k=0.01$ Mpc$^{-1}$. Let us recall that in this model $\Phi=\Psi$, therefore $\phi_{\rm len}=\Psi$.
We note that at late-time the gravitational potential for M1, M3 and M4 is enhanced with respect to $\Lambda$CDM, on the contrary for M2  it is suppressed. The largest deviation occurs for M4 ($\sim 26 \%$ at present time, $z\sim 0$), the model with the largest value of $A^2$, as expected from eq.~\eqref{eq:SCGMu}.  
Following this line of thinking, we would expect the growing order in deviation from $\Lambda$CDM to be M2 $<$ M3 $<$ M1 $<$ M4, considering the values of $A$ presented in Table \ref{tab:Models}. However, this is not the case: M3 is slightly above M1. This is due to the $\lambda$ parameter, whose value is noticeably smaller for M3 than M1, which enters in the $Q_sc_s^2$ term in eq.~\eqref{eq:SCGMu}. This makes $\mu(M3)>\mu(M_1)$ for $z+1<10^3$. Furthermore, the gravitational potential of M2 is suppressed with respect to $\Lambda$CDM, contrary to the small enhancement expected from eq.~\eqref{eq:SCGMu}. This is due to a suppression of the matter density perturbation with respect to the standard model. We further discuss this feature in Sec.~\ref{Sec:MatterPS}. These modifications in the lensing potential are mirrored in the lensing angular power spectra as shown in Figure \ref{fig:SCGLenPS}. Finally,  M4 is the only case which generates a significant large deviation from $\Lambda$CDM at early time as confirmed  by  the evolution of $\Psi$. The latter is connected to the evolution of $y_1$ which is the dominant component at early-time (see Figure \ref{fig:SCGy1}). Models that show early-time modifications of gravity \citep{Lin:2018nxe,Benevento:2018xcu} are known to alter the amplitude and phase of the high-$\ell$ acoustic peaks of the CMB temperature-temperature (TT) power spectrum. This is indeed what we find in Figure \ref{fig:SCGCMBTTPS} for M4.

\subsubsection{Impact on the Integrated Sachs-Wolfe (ISW) effect}

\begin{figure*}[t!]
\centering
\includegraphics[scale=0.35]{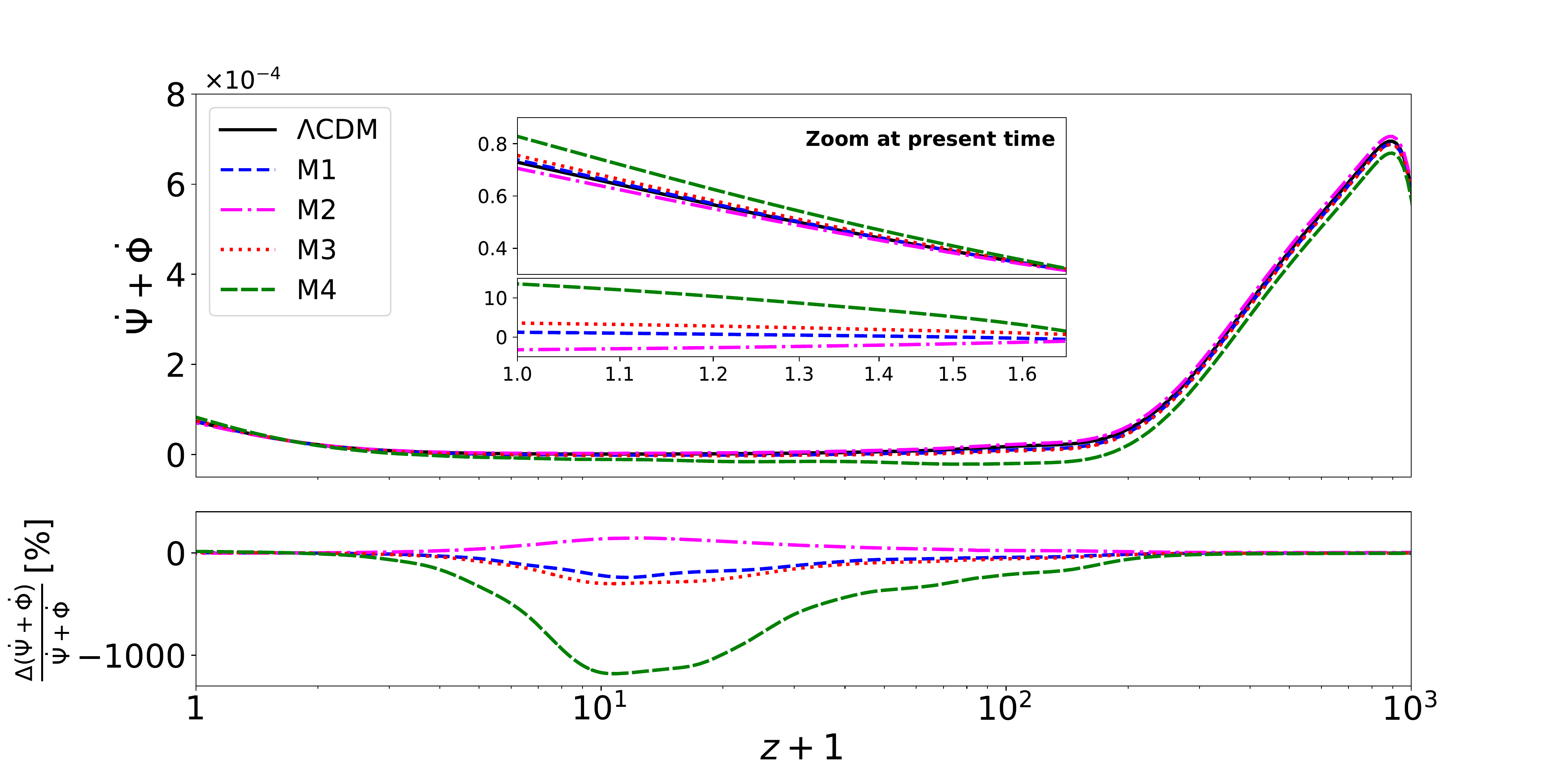} 
\caption[Evolution of the ISW source for the M models]{(Top) Evolution of the time derivative of $\Psi + \Phi$, computed at $k = 0.01$ Mpc$^{-1}$, for $\Lambda$CDM (solid black line) and the four M models of Table \ref{tab:Models}, namely: M1 (dashed blue line), M2 (dot-dashed magenta line), M3 (dotted red line) and M4 (long dashed green line). It is accompanied by a zoom at late-time of the same quantity. (Bottom) Percentage relative difference of the M models' $\Dot{\Psi} + \Dot{\Phi}$ computed with respect to $\Lambda$CDM.  \label{fig:SCGISWSour}}
\end{figure*}

A difference in the evolution of the gravitational potentials relative to the standard cosmological model impacts the ISW effect which is sourced by the time derivative of $\Psi + \Phi$. We show the evolution with redshift of $\dot{\Psi}+\dot{\Phi}$ for a fixed $k=0.01$ Mpc$^{-1}$ in Figure \ref{fig:SCGISWSour}. In the latter, M1, M2 and M3 closely follow the behaviour of $\Lambda$CDM for almost all $z$, showing some larger deviations at intermediate redshift ($10<z<100$). Then, while models M1 and M3 become slightly enhanced at present time, with the enhancements being of about $\sim 0.3\%$ and $\sim 2\%$ respectively, M2 is suppressed by $\sim 5\%$. M4 is the model which shows the larger deviations from $\Lambda$CDM at all redshift. Despite being suppressed with respect to the standard model for most of its evolution, it then becomes enhanced at late-time, reaching a $\sim 12\%$ enhancement relative to $\Lambda$CDM at present time.  

%
The change in the ISW effect affects the CMB TT angular power spectrum through the radiation transfer function \cite{Seljak:1996is}. We show the TT power spectra for the SCG and $\Lambda$CDM in Figure \ref{fig:SCGCMBTTPS}. Firstly, modifications in the time evolution of the gravitational potentials at late-time induce a late-time ISW effect which alters the low-$\ell$ tail. For the SCG,  M1, M3 and M4 show a suppressed ISW tail with respect to $\Lambda$CDM while M2 is slightly enhanced. Secondly, changes in $\dot{\Phi}+\dot{\Psi}$ during the transition from the radiation era to the matter one generate an early-time ISW effect that alters the amplitude of the first acoustic peak. Indeed we notice that M4's has a smaller amplitude. It has been found that models with a suppressed ISW tail are statistically favored by CMB data \cite{Frusciante:2019puu,Peirone:2019aua,Atayde:2021pgb}. On the contrary those with a large deviation in the late-time ISW source are ruled out \cite{Renk:2017rzu,Peirone:2017vcq}.

\begin{figure}[t!]
\centering
\includegraphics[scale=0.40]{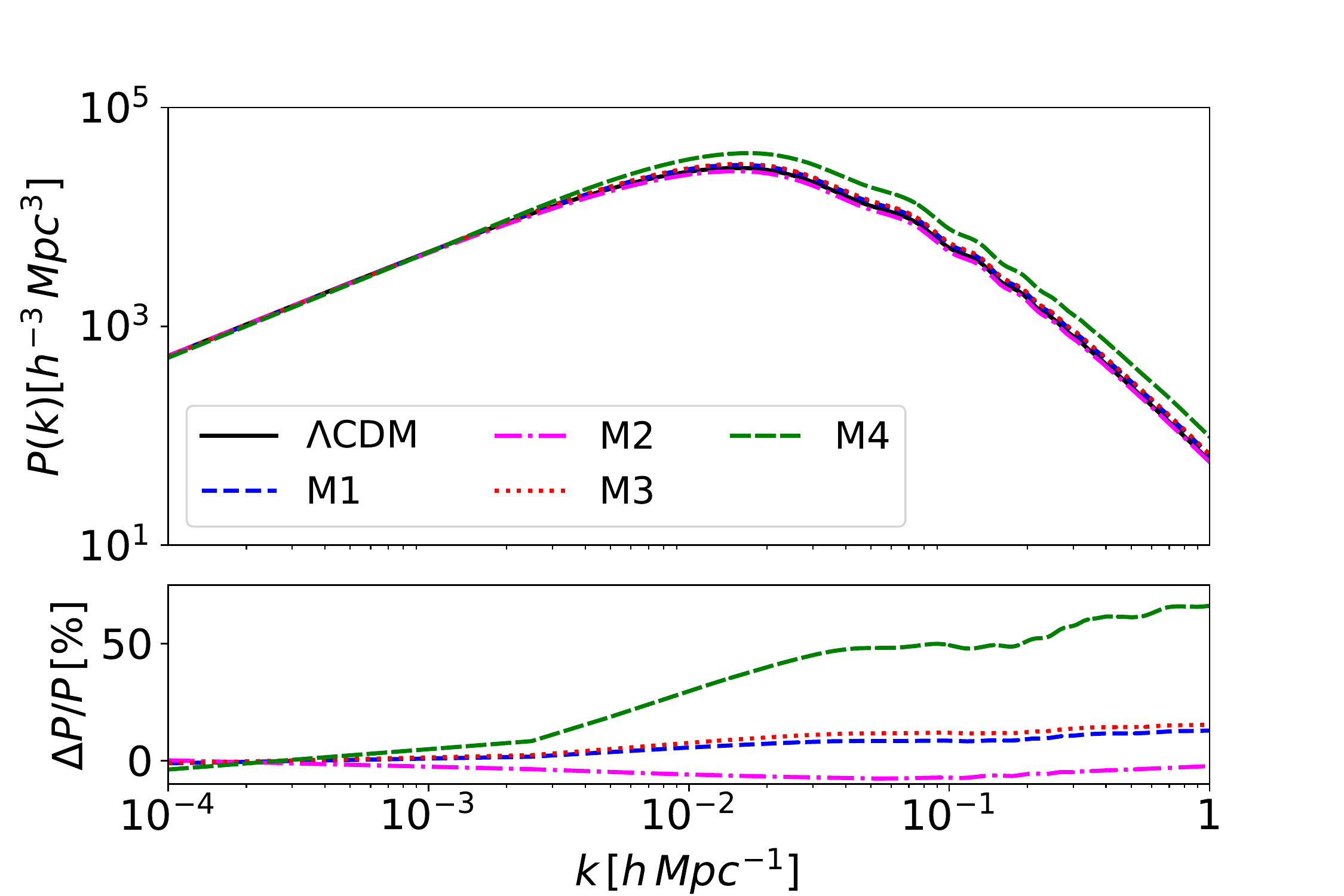} 
\caption[Matter Power Spectra for the M models]{(Top) Matter power spectra $P(k)$ for $\Lambda$CDM (solid black line) and the four M models of Table \ref{tab:Models}, specifically: M1 (dashed blue line), M2 (dot-dashed magenta line), M3 (dotted red line) and M4 (long dashed green line). (Bottom) Percentage relative difference of the matter power spectra computed with respect to $\Lambda$CDM. \label{fig:SCGMattPS}}
\end{figure}

\subsubsection{Impact on the growth of matter perturbations and the distribution of large-scale structures} \label{Sec:MatterPS}

The power spectrum of matter density fluctuations $P(k)$ for the SCG model shows deviations with respect to $\Lambda$CDM for $k > 10^{-3} \ h$Mpc$^{-1}$, as depicted in Figure \ref{fig:SCGMattPS}. We find it to be enhanced for models M1, M3 and M4 whereas M2 becomes suppressed. The enhancements are in agreement with the observed behaviour of $\mu$, with the largest deviation (up to $66 \%$ for larger $k$), being for the case with the largest value of $A^2$: M4.
%
\begin{figure}[t!]
\centering
\includegraphics[scale=0.4]{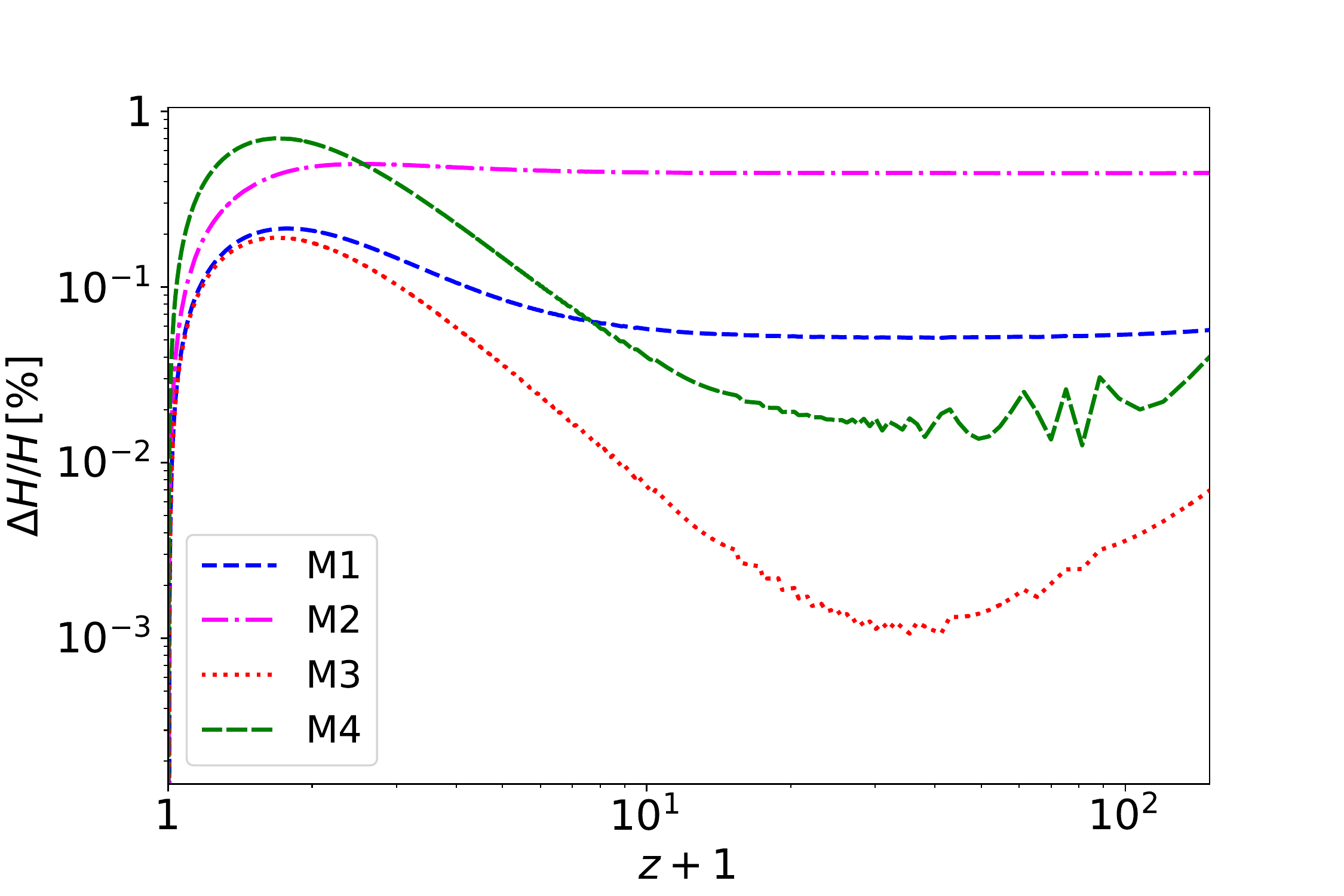} 
\caption[Difference in the Hubble rate]{Evolution of the relative difference of the Hubble parameter $H(t)$ for the M models featured in Table \ref{tab:Models} with respect to $\Lambda$CDM. Model M1 is presented in a dashed blue line, M2 in a dot-dashed magenta line, M3 in a dotted red line and M4 in a long dashed green line. \label{fig:SCGHDiff}}
\end{figure}
%
The suppression of M2 can be explained by considering that this case has the largest deviation in $H(t)$ relative to $\Lambda$CDM as shown in Figure \ref{fig:SCGHDiff}. This affects the evolution of the matter density perturbations because it enters as a friction term in the equation $\Ddot{\delta} + 2 H \Dot{\delta} -4 \pi G_N \mu (t) \rho \delta = 0$, hindering the growth of matter perturbations. The modifications introduced by $\mu$ are negligible for M2 because $\mu \sim 1$ at all times. 
The described effect has previously been observed for standard Quintessence \cite{Alimi:2009zk}.

\section{Observational Constraints} \label{Sec:Constraints}

Now that we have investigated the phenomenology of the SCG, we want to test the theoretical predictions with observations. In order to do this, we make use of \texttt{EFTCosmoMC} \cite{Raveri:2014cka}, a modification of the public \texttt{CosmoMC} software \cite{Lewis:2002ah,Lewis:2013hha}, that allows to sample the free parameter space using a  MCMC method, compute theoretical predictions through our modified version of \texttt{EFTCAMB} and compare them with observational data, thus reconstructing the posterior distribution of the sampled parameters.

As our baseline dataset we use here the {\it Planck} 2018 \cite{Planck:2019nip} (hereafter ''Planck'') data of CMB temperature likelihood for large angular scales $\ell =[2,29]$ and for the small angular scales a joint of TT, TE and EE likelihoods ($\ell =[30,2508]$ for TT power spectrum, $\ell =[30,1996]$ for TE cross-correlation and EE power spectra). 
 We also explore the effect on the constraints obtained adding other datasets to our baseline: we consider the combination Planck+lensing, where we also include the CMB lensing potential data from {\it Planck} ~\cite{Planck:2019nip,Planck:2018lbu}, and the combination Planck+BAO+SN where we also include baryon acoustic oscillations (BAO) data from the 6dF Galaxy Survey \cite{Beutler:2011hx}, the Sloan Digital Sky Survey (SDSS) DR7 Main Galaxy Sample \cite{Ross:2014qpa} and SDSS DR12 consensus release \cite{BOSS:2016wmc}, and supernova (SN) data from Pantheon \cite{Pan-STARRS1:2017jku}.

For all these combinations, our set of free parameters includes the standard cosmological parameters, i.e.\ the baryon and cold dark matter energy densities $\Omega_{\rm b}h^2$ and $\Omega_{\rm c}h^2$, the amplitude and tilt of the primordial power spectrum $A_{\rm s}$ and $n_s$, the optical depth $\tau$ and the angular size of the sound horizon at recombination $\theta_s$.  In addition to these, we also consider as free the SCG parameters $A$ and $\lambda$ and  we sample them in the range $A\in [0,3]$ and $\lambda \in [-100,0]$. Notice that here we consider the parameters $\beta_1$ and $\beta_2$ as fixed, given that, as it is discussed in Sec.~\ref{sec:CosmPert}, they have a negligible or no-impact on the observables under consideration.  We set them to the values $\beta_1=100$ and $\beta_2=0.7$. We nevertheless verified, in our baseline data case, that including them in the analysis and marginalising them out does not affect the final results. We use flat priors on all the sampled parameters.  Finally, we impose the stability conditions to avoid ghost and gradient instabilities \cite{DeFelice:2016ucp,Frusciante:2018vht} which are directly computed by \texttt{EFTCAMB} thanks to a stability built-in module, thus rejecting all sampled points in the parameter space that do not satisfy these conditions.

Once we obtain the sampled chains from \texttt{EFTCosmoMC} we analyze them using \texttt{GetDist} \cite{Lewis:2019xzd}.

\subsection{Results}

\begin{figure}[t!]
	\centering
	\includegraphics[width=1.\columnwidth]{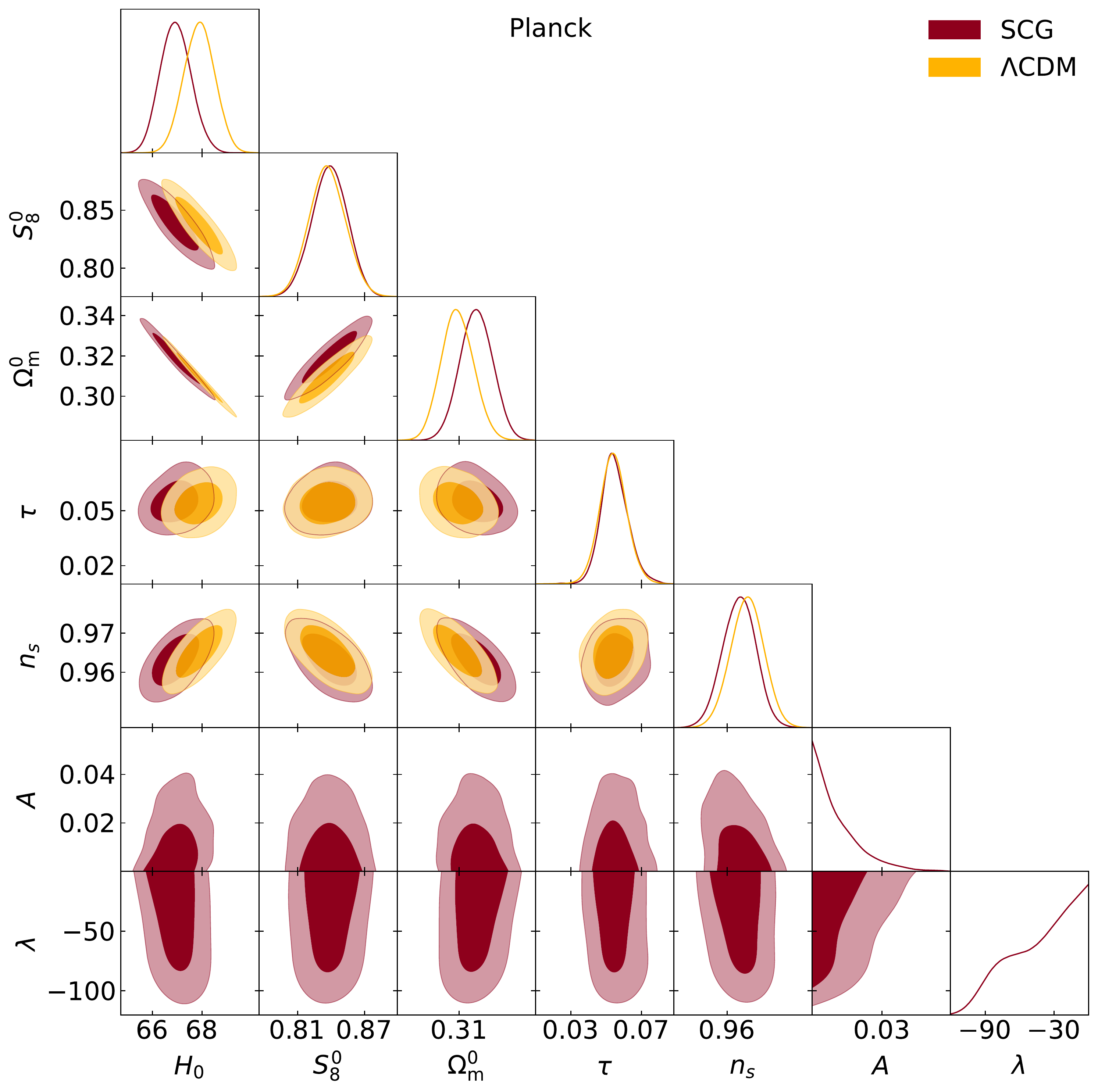}
	\caption{$68\%$ and $95\%$ confidence level contours obtained using the Planck dataset for the SCG model (red) and for a $\Lambda$CDM cosmology. }\label{fig:baseline}
\end{figure}

We show in \autoref{fig:baseline} and \autoref{tab:baseline} the results obtained in our baseline case, i.e.\ when only Planck data are considered, on the primary sampled parameters $\tau$, $n_s$, $A$ and $\lambda$, and on the derived parameters $H_0$, $\Omega^0_{\rm m}$ and $S^0_8=\sigma_8\sqrt{\Omega^0_{\rm m}/0.3}$, with $\sigma_8$ the dispersion of density perturbations on a scale of $8$ $h^{-1}$ Mpc.

We notice that the SCG parameter $A$, which can be considered as an amplitude of the deviation from GR, is constrained to be very small, thus highlighting the preference of Planck data for a GR cosmology. The parameter $\lambda$ is instead poorly constrained, mainly due to the fact that its effects become negligible as $A$ gets close to zero. 

Concerning the standard cosmological parameters, we notice that allowing for a SCG cosmology has the effect of shifting the constraints on $H_0$ towards smaller values with respect to the $\Lambda$CDM case, while $\Omega^0_{\rm m}$ takes slightly larger values and the other parameters are mostly unchanged. The first of these effects however is significant; the Planck data give a value for $H_0$ in $\Lambda$CDM that is in tension of approximately $4.5\sigma$ with local measures (see e.g.\ \cite{DiValentino:2020zio} for an extensive review), which prefer higher values of this parameter. While one hopes that new physics would be able to reconcile this tension, we notice that the SCG does not allow to ease the difference between low and high redshift observables, as it instead increases it even if not in a significant way. In Quintessence models it has been found that having $w_{\phi} > -1$ worsens the $H_0$ tension \cite{Banerjee:2020xcn}. For our model, if we consider the mean values obtained with the Planck dataset we find $w_{\phi,0} = - 0.98$. As for Quintessence, this might be one of the reasons why the SCG is incapable of solving the $H_0$ tension.

In addition to this, we notice that the minimum $\chi^2$ ($\chi^2_{\rm min}$) increases when one fits the Planck data in the SCG model with respect to the $\Lambda$CDM model. This appears to be counter intuitive, as the former model has two additional parameters with respect to the latter; however, the SCG model we investigate does not have a $\Lambda$CDM limit, but rather reduces to Quintessence when $A=0$, a limit that anyway lies at the very edge of the prior range. For such reason, we can interpret this increase in $\chi^2_{\rm min}$ as an hint that $\Lambda$CDM is still the model favoured by the data, with even the Quintessence $A=0$ limit of SCG having a worse goodness of fit with respect to it.

\begin{table}[t!]
    \centering
    \begin{tabular}{c|c|c}
Parameter          &  SCG                          & $\Lambda$CDM\\
\hline
$\tau$             & $0.0548^{+0.0068}_{-0.0080}$ & $0.0540\pm 0.0078$\\
$H_0$              & $66.94\pm 0.61$              & $67.88\pm 0.62$\\
$\Omega^0_{\rm m}$ & $0.3189\pm 0.0083$           & $0.3091\pm 0.0083$\\
$S^0_8$            & $0.839\pm 0.016$             & $0.837\pm 0.016$\\
$n_s$              & $0.9632\pm 0.0044$           & $0.9653\pm 0.0044$\\
$A$                & $< 0.0133$                   & ...\\
$\lambda$          & $> -56.0$                    & ...\\
\hline
$\chi^2_{\rm min}$ & $2774$                       & $2771$
    \end{tabular}
    \caption{$68\%$ limits on the cosmological and SCG parameters obtained using Planck data for a SCG and a $\Lambda$CDM cosmology and their  minimum $\chi^2$ ($\chi^2_{\rm min}$).}
    \label{tab:baseline}
\end{table}

\begin{figure}[!h]
	\centering
	\includegraphics[width=1.\columnwidth]{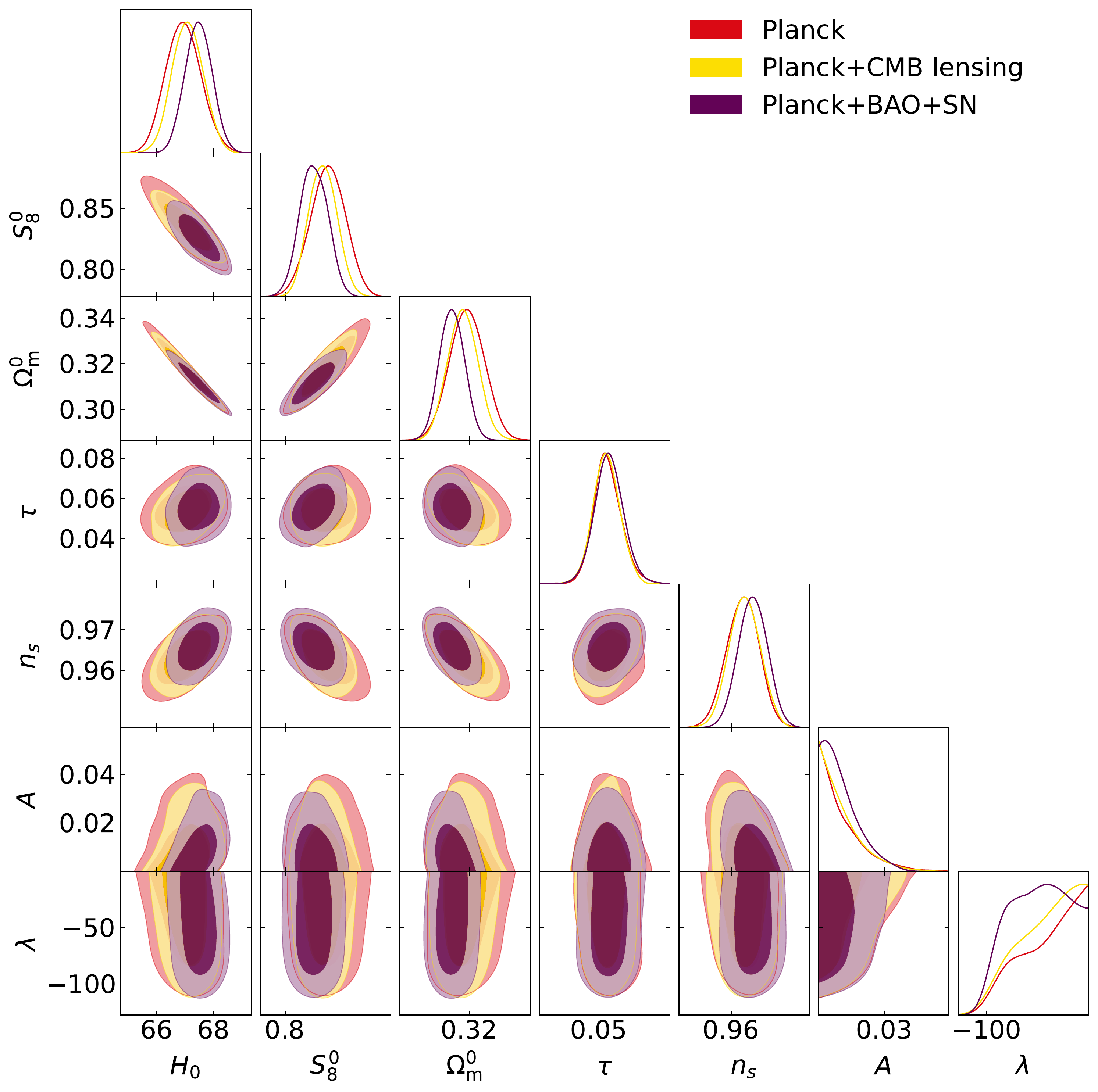}
	\caption{$68\%$ and $95\%$ confidence level contours obtained in a SCG cosmology for the Planck data (red), the Planck+lensing combination (yellow) and Planck+BAO+SN (purple).\label{fig:allobs}}
\end{figure}

In \autoref{fig:allobs} and \autoref{tab:allobs} we compare instead our baseline results on the SCG model with those obtained when using the data combinations Planck+lensing and Planck+BAO+SN.
We notice that the inclusion of additional dataset does not improve significantly the constraints on the SCG parameters, with the exception of Planck+BAO+SN that seems to exhibit a peak for $\lambda$ for a non-vanishing value. However, despite a good convergence of the chains, this effect might be due to the aforementioned degeneracies between $\lambda$ and $A$ with the former that is allowed to take any value when the latter vanishes, a degeneracy that might not allow a good sampling with the Metropolis-Hastings algorithm employed by EFTCosmoMC. 

The bounds on the cosmological parameters are instead tightened by the inclusion of the additional datasets, with the important features of $H_0$ being brought back to the Planck value for $\Lambda$CDM when we consider Planck+BAO+SN. In this case, the data still allow for a departure from GR, encoded in the non-vanishing value of $A$, avoiding the worsening of the tension with low redshift data that we noticed in the Planck case. 

\phantom{text needed to not mess up the layout}

\begin{table}[!h]
    \centering
    \begin{tabular}{c|c|c|c}
Parameter          &  Planck                      & Planck+lensing     & Planck+BAO+SN\\
\hline
$\tau$             & $0.0548^{+0.0068}_{-0.0080}$ & $0.0541\pm 0.0074$ & $0.0556\pm 0.0079$\\
$H_0$              & $66.94\pm 0.61$              & $67.08\pm 0.54$    & $67.46\pm 0.47$\\
$\Omega^0_{\rm m}$ & $0.3189\pm 0.0083$           & $0.3168\pm 0.0072$ & $0.3116\pm 0.0059$\\
$S^0_8$            & $0.839\pm 0.016$             & $0.834\pm 0.013$   & $0.826\pm 0.012$\\
$n_s$              & $0.9632\pm 0.0044$           & $0.9637\pm 0.0042$ & $0.9659\pm 0.0039$\\
$A$                & $< 0.0133$                   & $< 0.0126$         & $< 0.0129$\\
$\lambda$          & $> -56.0$                    & $> -55.9$          & $49^{+36}_{-27}$\\
    \end{tabular}
    \caption{$68\%$ limits on the cosmological and SCG parameters obtained in a SCG cosmology for the dataset combinations Planck, Planck+lensing and Planck+BAO+SN.}
    \label{tab:allobs}
\end{table}

\section{Conclusion}\label{Sec:Conclusion}

We studied for the first time the phenomenology at large linear scales of a Scaling Cubic Galileon model given by the Lagrangian \eqref{eq:SCGaction} and characterized by four additional parameters $\{A,\lambda, \beta_1, \beta_1\}$ in comparison to the standard cosmological model. Furthermore, we provided the cosmological constraints using CMB, CMB lensing, BAO and SN data.

The model shows very interesting features. Firstly, the early time scaling regime after which solutions approach a late time attractor offers the model the opportunity to be compatible with particle physics' energy scale at early-time while still realizing late-time cosmic acceleration. Regarding the evolution of linear cosmological perturbations, the modifications of the gravitational potentials are identified to be weighed largely by the parameter $A$ even though $\lambda$ can also have a relevant impact. 
The remaining two parameters, $\beta_1$ and $\beta_2$, have no direct impact on the perturbations but they change the parameter space of $A$. 
We identify three main effects on the cosmological observables: deviations in the lensing potential, which can be either suppressed or enhanced with respect to the $\Lambda$CDM model and correspondingly leads to a suppressed/enhanced lensing angular power spectrum. Additionally, the changes in the lensing potential also modify the high-$\ell$ acoustic peaks of the temperature-temperature power spectrum. Following this, a change in the time derivative of the gravitational potentials is then expected which modifies the ISW effect. Indeed we found that modifications in the early ISW effect led to a lower amplitude of the first acoustic peak of the temperature-temperature power spectrum for the SCG model. The latter is also modified in the low-$\ell$ tail where in this case the differences with respect to $\Lambda$CDM were due to modifications in the late ISW effect. Finally, the power  spectrum  of  matter  density  fluctuations is also enhanced/suppressed with respect to the standard scenario. The modifications are driven mostly by $A^2$ which enhances the matter power spectrum. We found that there are cases for which  modifications in the background expansion give rise to a friction term that dominates over any other modified gravity source, leading to a suppressed matter power spectrum.

We tested these effects with cosmological data and we found that the SCG model is also suffering from the $H_0$ tension when using only Planck CMB data which is even worse than the $\Lambda$CDM model. The joint analysis of CMB, BAO and SN data was able to set an upper bound on the parameter $A<0.0129$ and to constrain $\lambda=49^{+36}_{-27}$ at 68\% C.L.. The parameter $A$ is then very close to zero which in turn led to loose power in constraining on $\lambda$. Furthermore, the SCG model exasperates the $H_0$ tension between Planck data and local measurements.

In conclusion if on one side the SCG model eases some issues of the $\Lambda$CDM model, on the other side the $H_0$ tension is still present. It would then be of interest to consider the model for further investigations in the future, particularly once new data are available, which will also help in shedding light on the nature of this tension.

\begin{acknowledgements}
We thank N.~J.~Nunes for useful discussion. 
ISA and NF acknowledge support by Funda\c{c}\~{a}o para a  Ci\^{e}ncia e a Tecnologia (FCT) through the research grants UIDB/04434/2020, UIDP/04434/2020, PTDC/FIS-OUT/29048/2017, CERN/FIS-PAR/0037/2019, PTDC/FIS-AST/0054/2021. The research of ISA has received funding from the FCT PhD fellowship grant with ref.~number 2020.07237.BD. NF  acknowledges support from her FCT research grant ``CosmoTests -- Cosmological tests of gravity theories beyond General Relativity" with ref.~number CEECIND/00017/2018.
MM acknowledges support from ``la Caixa'' Foundation (ID 100010434), with fellowship code LCF/BQ/PI19/11690015, and the Centro de Excelencia Severo Ochoa Program SEV-2016-059. 
\end{acknowledgements}

\normalem
\bibliography{Bib_entries}

\end{document}